\documentclass[useAMS,usenatbib]{mn2e}
\usepackage{graphicx}
\usepackage{latexsym}
\usepackage{amsmath}
\usepackage{subfigure}
\title[Photoevaporation of Discs]{The imprint of photoevaporation on edge-on discs}
\author[J. E. Owen, B. Ercolano \&
    C. J. Clarke]{James E. Owen$^1$\thanks{E-mail: jo276@ast.cam.ac.uk},
    Barbara Ercolano$^{2}$ \& Cathie J. Clarke$^{1}$ \\ $^1$Institute of
    Astronomy, Madingley Road, Cambridge, CB3 0HA, UK\\ $^2$School of
    Physics, University of Exeter, Stocker Road, Exeter EX4 4QL}

\begin{document}

\pagerange{\pageref{firstpage}--\pageref{lastpage}} \pubyear{2002}

\newcommand{\dd}{\textrm{d}}
\newcommand{\rin}{R_\textrm{\tiny{in}}}
\newcommand{\OO}{\mathcal{O}}
\maketitle

\label{firstpage}

\def\mnras{MNRAS}
\def\apj{ApJ}
\def\aap{A\&A}
\def\apjl{ApJL}
\def\apjs{ApJS}
\def\araa{ARA\&A}

\begin{abstract}
We have performed hydrodynamic and radiative transfer calculations of a
    photoevaporating disc around a Herbig Ae/Be star to determine the
    evolution and observational impact of dust entrained in the
    wind. We find that the wind selectively entrains grains of
    different sizes at different radii resulting in a dust population
    that varies spatially and increases with height above the disc at
    radii $>10$ AU. This variable grain population results in a
    `wingnut' morphology to the dust density distribution. We calculate
    images of this dust distribution at NIR wavelengths that also show
    a wingnut morphology at all wavelengths considered. We have also
    considered the contribution that crystalline dust grains  will have
    in the wind and show that a photoevaporative wind can result in a significant crystallinity fraction at all radii, when the disc is edge-on. However, when the disc's photosphere is unobscured, a photoevaporative wind makes no contribution to the observable crystallinity fraction in the disc. Finally, we
    conclude that the analysis of extended emission around
    edge-on discs could provide a new and independent method of testing
    photoevaporation models.   
\end{abstract}

\begin{keywords}
accretion, accretion discs - circumstellar matter - planetary systems:~protoplanetary discs - stars:~pre-main-sequence.
\end{keywords}

\section{Introduction} 
Protoplanetary discs are ubiquitous around young stars, an inevitable
    consequence of angular momentum conservation during star
    formation. Young stars accrete their mass through discs  which then
    provide the reservoir of material from which planets may form. Any
    understanding of the processes involved in star and planet
    formation requires knowledge of how discs evolve and are finally
    dispersed. Observations of disc evolution have typically focused
    on thermal emission from the dust grains in the disc
    through the analysis of their continuum spectral energy
    distribution (SED) at several different 
    wavelengths. While empirical analysis and numerical modelling
      of the SED of discs around young stars have proved extremely
      useful to study this important phase of the star and planet
      formation process, a number of well known degeneracies conspire
      to render the interpretation of the data non-unique. To remedy
      these shortcomings recent observational efforts are making use
      of various techniques to obtain spatially resolved information
      of many sources. 
 
Herbig Ae/Be stars are more luminous than their lower mass
T-Tauri counterparts and provide an opportunity to observe large
    scale morphology in resolved images. Discs observed at
      nearly edge-on inclinations
    are particularly useful in studying the large scale structure
    since the star and inner disc, which are much brighter than any
    extended emission, are obscured along the line of sight. Several
    edge-on discs around young stars exhibit extended
    emission above and below their mid-plane at
    NIR wavelengths (e.g. Padgett et al. 1999 \& Perrin et al. 2006)
    and some have been further imaged through AO systems
    providing excellent spatial resolution, as in the case of PDS 144N
    (Perrin et al. 2006). 

Photoevaporation (Hollenbach et al. 1994)  is the mechanism by which
    high energy photons (UV-X-rays) heat the surface layers of the
    discs to temperatures of order the escape temperature, forming a
    thermally driven wind from the disc out to large
    radius. Photoevaporation has been proposed as the dominant disc
    dispersal mechanism around low-mass stars (e.g. Clarke et
    al. 2001) and has been successful in explaining many of the
    observational statistics (Owen et al. 2010b). However there is
    only tentative evidence that the photoevaporative wind itself has been
    detected, through NeII and OI emission (Hartigan et
    al. 1995; Alexander et
    al. 2008; Pascucci et al 2009; Schisano et al 2010; Ercolano \&
    Owen 2010) which 
    probes the wind on a very local scale. In order to test
    photoevaporation, along with the various different heating mechanisms we need
    global observational diagnostics to validate the models. Edge-on
    discs provide the perfect opportunity to detect a disc wind on a
    global scale through light scattered by the dust
      grains that will inevitably be entrained by the wind. 

    The origin of crystalline dust grains detected
    in protoplanetary discs via infra-red (IR) spectroscopy
    (e.g. Bouwman et al. 2001) and directly within our own solar
    system (e.g. Wooden et al. 1999) is still a matter of discussion. 
The source of all dust in star formation - the ISM - is inferred to be
entirely 
    amorphous (e.g. Kemper et al. 2005). Spectroscopy
    of discs (van Boekel et al. 2005; Apai et al. 2005) indicates
    a non-negligible level of crystalline dust 
    outside the crystalline radius, where grains are hot
    enough ($T>800$K) to be thermally annealed and converted into crystalline
    grains. This discovery has lead to the development
    of disc models with radial mixing to allow 
    crystalline grains that formed in the hot inner disc to be transported to larger radii
    (e.g. Morfill \& Voelk 1984; Gail 2001, 2002; Wehrstedt \& Gail
    2002; Bockelee-Morvan et al. 2002; Dullemond et
    al. 2006; Hughes \& Armitage 2010). Another suggestion put forward by Shu et al. (1996), is
    that crystalline grains could be carried outwards by a x-wind. Given that the crystallization radius of $<$1-2AU is
    smaller than the minimum launch radius for a photoevaporative wind
    $>2AU$ around Herbig Ae/Be stars and lower mass stars, the photoevaporative wind cannot
    be the direct source of the crystalline grains at larger
    radii, but could work in combination with radial mixing to produce
    the observed enhancement. Furthermore crystallinity around edge-on discs
    can be used to probe the source of the dust in the extended emission, since a
    disc origin will give rise to crystalline dust grains in the wind
    while an in-falling envelope will contain only amorphous grains.
    
    In this paper we present a simple model to investigate the
    imprint of
    photoevaporation on the extended
    emission observed around edge-on Herbig Ae/Be stars. We will
    compare our model to current observations of edge-on discs along with
    following the fate of crystalline grains in the wind as they are
    transported to large radius. In Section~\ref{model} we
      describe the model, 
    including the hydrodynamic and radiative transfer methods. In
    Section~\ref{sec:images} we present synthetic images obtained from
    the models. In Section~\ref{sec:cryst} we describe the
    results of crystallinity calculations for the wind. We
    compare our results to observations of edge-on discs in
    Section~\ref{sec:compare} and we summarise our main findings in
    Section~\ref{sec:conclusions}.   
    
\section{Model}\label{model}
Considering the 
    force balance on a dust grain, it is easy to show that small dust
    particles are entrained in a photoevaporative wind, which carries
    them out to large distances.
Takeuchi et al. (2005) showed that the drag
    force on a grain is approximately:
    \begin{equation}
      F_d\approx\frac{m_d\rho_wv_w^2}{\rho_da}
    \end{equation}
\noindent where $m_d$, $\rho_d$ and  $a$ are, respectively; the mass, the
   density and the radius of a dust grain (assumed to be spherical) and
$\rho_w$ and $v_w$ are the density and velocity of the wind. At
large radius \footnote{Throughout this work we use ${R,\phi,z}$ to refer to
cylindrical co-ordinates and ${r,\theta,\varphi}$ to refer to spherical polar co-ordinates}($z/R>1$) the flow is approximately spherical implying 
    $\rho_wv_w$ falls  off as $1/r^2$. Since gravity also falls off as
    $1/r^2$ and $v_w$ increases monotonically with radius, then if a
    grain is still entrained several scale heights above the launching
    surface, it will be entrained permanently, allowing dust to be
    carried to very large distance from the star, since the drag force
    will dominate over gravity. 

   We build a simplified model to test the
    possibility of extended emission from dust grains due to a
    photoevaporative wind, and we choose a set of assumptions that
    allows us to place an upper limit on this expected level of emission. Namely
    we assume that the 
    entire dust population is able to reach the launching surface of
    the wind via some turbulent mechanism and ignore the effects
    of settling and grain growth. Our simple model ignores the
  details of the bound regions of the disc\footnote{In the sense
    that $T\ll T_{\textrm{escape}}$} and thus ignores any emission
  produced by this region,  including {\it only} 
    the wind itself and its contributions to emission. This simplification means our model will not be accurate near the mid-plane, where emission from dust in the disc's upper atmosphere dominates over the wind emission. Therefore, we are unable to reproduce the observed intensities and optical depths seen at several scale height above and below the mid-plane in an edge-on disc; however, here we are interested in the extended emission, which, in the absence of an infalling envelope is dominated by the photoevaporative wind.

Our model naturally produces a `wingnut' morphology similar to
that seen in several scattered light images (e.g. Perrin et
al. 2006), and produces a spatially variable 
    dust distribution, due to the different maximum grain
    size that can be entrained along each streamline. The radial morphology  of the streamlines then yields that, at a given cylindrical radius, the maximum grain size increases with height, resulting in a spatial variation of colour in the scattered light images. Furthermore crystalline grains entrained in
    the wind  are 
    transported  outwards from  regions of the disc with
    higher crystallinity fractions resulting in an enhancement
    of crystalline grains in the wind over the disc's underlying distrubtion.  

We can
    split the method for constructing scattered light images into
    three separate parts: (i) hydrodynamic calculations of the
    photoevaporative wind;  (ii) calculation of the dust profile
    distribution and crystallinity based on the hydrodynamic
    solution; (iii) radiative transfer modeling of the dust distribution. 

\subsection{Model Set}
As the mass-loss rates are not well known for
    Herbig Ae/Be stars we have left the ionizing luminosity (which
    sets the mass-loss rates) for an EUV driven wind (Hollenbach et
    al. 1994) as a free parameter and consider the
    effect of changing mass-loss rates on the morphology, colour
    of the emission and crystallinity distribution in the winds. 
We consider an EUV wind from a primordial disc around a 2.5M$_\odot$ star with a
    range in ionizing luminosity from $10^{41}$ to $10^{45}$ s$^{-1}$
    in steps of 1~dex. These luminosities correspond to mass-loss
    rates in the range $10^{-10}-10^{-8}$M$_\odot$yr$^{-1}$ and are
    similar to values calculated by Alexander et al. (2004). In order to
    calculate images we compute scattered light images at 1.6, 2.1
    \& 3.8$\mu
    $m with band passes corresponding to the H, K' \& L' bands. In
    order to consider the relative contributions of crystalline to
    amorphous dust we take the calculated crystallinity fractions in
    the disc of
    Dullemond et al. (2006) for an evolved Herbig Ae/Be star. We
    assume that this disc crystallinity fraction is the same at the
    base of our photoevaporative wind, such that the crystallinity
    fraction along the streamline can be calculated by following the
    thermal evolution of the dust entrained on that streamline. 

\subsection{Hydrodynamic photoevaporative wind}
A photoevaporative disc wind is a thermally driven hydrodynamic wind
occuring when the disc's surface is heated to  temperatures of
order the escape  
    temperature, allowing it to launch a freely expanding wind. While
     the wind driving source for lower mass (T-Tauri) stars is likely to be X-rays
    (Owen et al. 2010a, Ercolano \& Owen 2010, Ercolano \& Clarke
      2010, Owen et al. 2010b) the  wind driving source around
    intermediate mass stars has not yet been  thouroughly
    investigated. X-ray photoevaporation may still occur to some
    degree; however, the lower $L_X/L_{bol}$ ratio of Herbig Ae/Be
    compared to T-Tauri stars and their higher EUV fluxes, may
    reduce the role of X-rays in driving the wind. 
 We adopt here the EUV  driven wind of Hollenbach et
    al (1994) and hydrodynamic solution of  Font et al. (2004), which
    is a {\it simple} and {\it scalable} hydrodynamic
    solution, allowing us to consider a wide range of parameter
    space, something not possible  with more complicated FUV models (Gorti \&
    Hollenbach 2009) or
    X-ray models (Owen et al. 2010a).  The simplified EUV treatment is suitable for
      the purpose of this work which aims at being the first approach
    in studying the qualitative
      aspects of scattered light emission from a disc wind. We have
    repeated the calculation of Font et al. (2004) and Alexander (2008)
    and we refer the reader to these papers for a
    detailed description of the model setup, while the basics are
    summarised below.
    \subsubsection{Numerical EUV Photoevaporative Wind}
\begin{figure}
\centering
\includegraphics[width=\columnwidth]{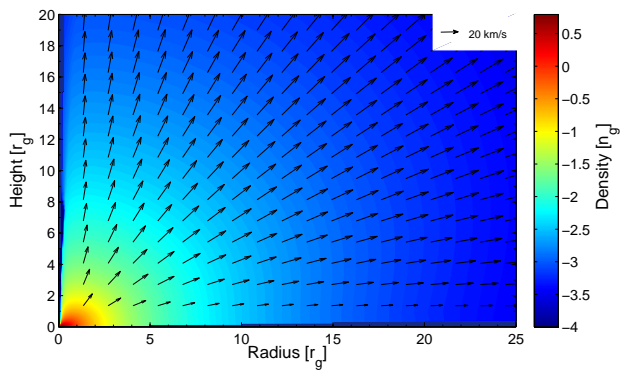}
\caption{EUV wind structure: the colour-map shows the
    density, while the arrows show the direction and magnitude of the
    gas flow, note that just after launch the flow is very close to
    being spherical.}\label{fig:structure}
\end{figure}

    In order to determine an accurate kinematic structure of the wind
    we must compute a numerical solution to the problem. We use the
    {\sc zeus2d} code (Stone \& Norman 1992); we employ a spherical grid with
    $\theta=[0,\pi/2]$, the radial grid cells are logarithmically
    spaced, such that we have adequate resolution at small radius to
    resolve the onset of the flow. The calculation is an isothermal
    wind calculation with the sound speed set to $c_s=10$km s$^{-1}$
    and the radius scaled to the length scale $r_g=GM_*/c_s^2$
    i.e. the radius at which the internal energy of the gas is enough
    to unbind the gas from the star. We use a radial range of
    $r=[0.05,40]r_g$ with $N_r=240$ and $N_{\theta}=50$. The number
    density at the  base (i.e. the density along the $\theta=\pi/2$ axis) of the wind was calculated semi-analytically
    by Hollenbach et al. (1994) and scales as $R^{-3/2}$ for $R<R_g$ and
    as $R^{-5/2}$ outside $R_g$, as in Font et al. (2004) and
    Alexander et al. (2008) we
    adopt the smooth {\bf base } density profile suggested by \citet{font04} that
    varies between the two power laws:
\begin{equation}
n(R)=n_g\left[\frac{2}{(R/R_g)^{15/2}+(R/R_g)^{25/2}}\right]^{1/5}
\end{equation}
\noindent where $n_g$ is the density at $R_g$ which was determined through the
    numerical calculations of \citet{hollenbach94} to be:
\begin{equation}
n_g\approx2.8\times
    10^4\left(\frac{\Phi}{10^{41}\textrm{s}^{-1}}\right)^{1/2}\left(\frac{M_*}{1\textrm{M}_\odot}\right)^{-3/2}\textrm{cm}^{-3}
\end{equation}
 where $\Phi$ is the ionizing luminosity. We note that this hydrodynamic calculation does not include a cold bound `disc' component, since as discussed above the base density structure of the wind is known a priori.   This base density, along with Keplerian rotation is  reset at every
    time-step and the model is allowed to evolve to a steady state launching a wind from the grid's mid-plane (representing the disc's surface). As
    expected, we find excellent agreement with the results of the
    \citet{font04} and \citet{alexander08} calculations. In
    Figure~\ref{fig:structure} we show a plot of the converged wind
    structure showing that the wind is approximately spherical once it
    has reached several scale heights, in agreement with our earlier
    discussion in Section~\ref{model}.

\subsection{Dust Distribution}\label{sec:dust}
\begin{figure}
      \centering
      \includegraphics[width=\columnwidth]{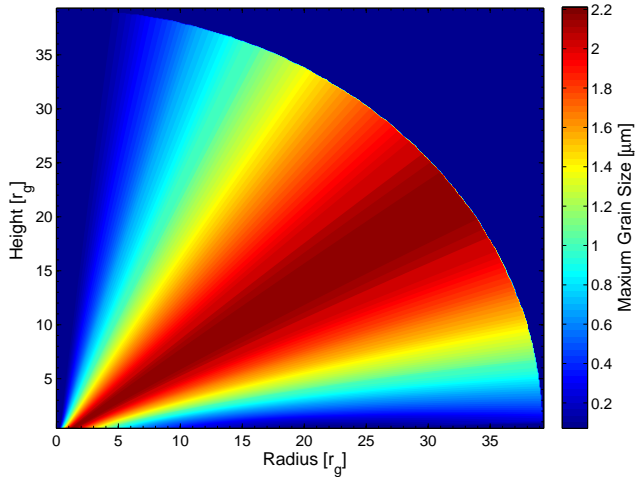}
      \caption{Colour map of the maximum size of entrained dust
    particles at a given position for $\Phi=10^{43}$s$^{-1}$, we note that most dust comes
    from within 1-5$r_g$.}
\label{fig:amax}
\end{figure}

In order to calculate the dust distribution in the wind, we must make
    some assumptions about the underlying dust distribution in the
    disc. We adopt a dust to gas mass ratio of 0.01, and a power law
    grain size distribution of index -3.5 (Mathis, Rumpl \& Nordsiek,
    1977, MRN), with grain sizes ranging from
    $a_{min}$=5$\times10^{-3}\mu$m to $a_{max}$=1mm, we assume
    spherical grains with a density of 1g~cm$^{-3}$. We also
    assume that the dust  is fully mixed within the disc up to
    the transition between the bound cold disc and the hot
    EUV heated flow. We then calculate streamlines from the base of
    the flow to the edge of the grid, then along each streamline,
    compute the force balance between the drag force (calculated from
    Equation 1), gravity and the centrifugal force. We take the
    dust grain as entrained if the net force along the streamline is
    $>0$. We then obtain the maximum grain size entrained along the
    entire streamline (making sure to check that each grain can be
    entrained the entire length of the streamline). In
    Figure~\ref{fig:amax} we show the obtained maximum grain size as a
    function of position in the flow for
    an ionizing luminosity of $\Phi=10^{43}$s$^{-1}$. 

\begin{figure}
\centering
\includegraphics[width=\columnwidth]{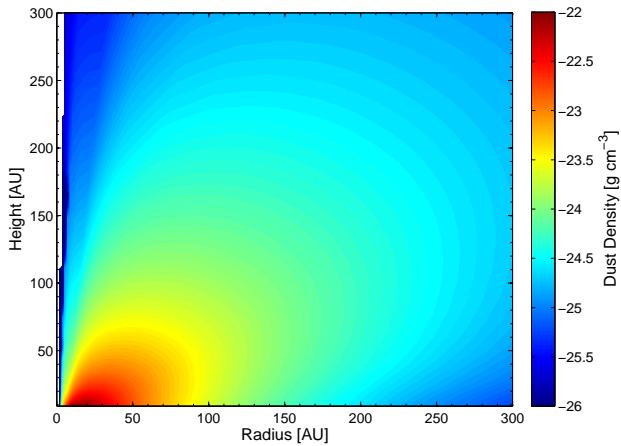}
\caption{Dust mass density distribution in the wind model for the
    $\Phi=10^{43}$s$^{-1}$ calculation.}
\label{fig:ddensity}
\end{figure}

We can
    then compute the dust density from this maximum grain size
    under the assumption that the dust all comes from the same
    underlying population described above. In
    Figure~\ref{fig:ddensity}
we show the obtained dust density, noting that the combination of a
    photoevaporative wind and a selective dust population naturally reproduces a `wingnut'
    structure.

\subsection{Radiative Transfer}\label{sec:rad}
We calculate the radiative transfer as a two step process: First we calculate the
    temperature structure of the dust using a 3D grid calculated from
    the 2D dust density calculations. We
    have modified the {\sc mocassin} code (Ercolano et al. 2003, 2005,
    2008) to allow cell-to-cell variations of the grain size
    distribution. We interpolate the
    results from Section~\ref{sec:dust} onto at 3D Cartesian grid with
    resolution $N_X\times N_Y\times N_Z=99\times 99\times50$ assuming
    azimuthal symmetry, using the range in the x and y axis of [-32,32]$r_g$
    and in the z axis of [0,32]$r_g$. 
   
     The input SED used to irradiate the wind is taken from the model
    set of Robitaille et al. (2006), Model Number:3011099, 
    corresponding to a 2.51M$_\odot$ star with an effective
    temperature of 9480K and radius of 2.18R$_\odot$, surrounded by an
    optically thick disc extending into the dust destruction
    radius. The irradiating SED is placed at the origin of the
     Cartesian grid; this
    approximation is valid in this case where we are mainly interested
    in scattering of near-IR (NIR) photons on the much larger scale of the
    wind i.e. on scale much greater than the 
    scale the disc emission arises from. 

In order to calculate images of the dust distribution, we have
    written a simple ray tracing driver for {\sc mocassin}, which we
    have called {\sc mocassinThinImage}. The standard
    Monte-Carlo radiative transfer techniques used to calculate
    scattered light images
    in {\sc mocassin} (i.e. by capturing  scattered photons along the line
    of sight of interest) is inappropriate for use in this case. The wind
    structure is too optically thin to obtain a converged image with a
    computationally feasible number of photon packets, since most packets
    leave the grid without interacting with the
    distribution\footnote{It is important to emphases this is not the
    case for the temperature calculation discussed above, where an interaction is not
    required with a photon packet to include its contribution to the
    radiative energy density and hence temperature in that cell (see
    Lucy 1999; Ercolano et al 2003).}. {\sc
    mocassinThinImage} performs a ray tracing calculation throughout
    the grid in both scattered and thermal emission, under the
    assumption that the medium is optically thin 
    (true of the wind always). We note again that since we have not
    explicitly included the dust in the `bound' region of the disc we
    will not calculate scattered light emission from this
    region. We also assume the disc mid-plane is optically thick to all
    radiation and hence only consider one hemisphere of the wind. 

\subsection{Calculations of Crystallinity Structure}\label{sec:mcryst}
In order to consider the structure of the dust grains in the wind, we have
    built a very simple dust crystallinity model. We assume that the
    grains are launched from a disc that is infinitesimally thin and
    has the crystalline 
    structure calculated by Dullemond et al. (2006), for a Herbig star
    with a mass of 2.5$M_\odot$ after 3.3Myr of evolution.  Any dust
    that was amorphous and is heated to a temperature $>800$K while
    entrained in the wind is then
    assumed to immediately become totally crystalline. Any dust that starts off
    crystalline or becomes crystalline at any point in the flow is
    assumed to be crystalline for the remainder of the flow regardless
    of the temperature it reaches at any other stage (we find that no
    dust in the flow reaches a temperature exceeding the
    evaporation temperature, which would destroy crystalline dust
    grains). In order to calculate a radial profile of the fraction of dust  in
    the wind that is crystalline/amorphous we must apply some cut off
    on the flow since the total mass of the flow is a function of time
    since it switched on, rather than a converged quantity (it diverges
    logarithmically). Since much of the information on dust structure
    comes from the 10$\mu$m silicate feature a temperature cut-off is
    most appropriate, thus we calculate the mass-fraction of
    crystalline grains out to a temperature of 100K; as such this
    should provide a reasonable estimate of the observable
    crystallinity-fraction (anything colder is extremely unlikely
    to make much impact on the 10$\mu$m silicate feature). Furthermore, since each cylindrical radial 
    region will be cut by many streamlines with different crystallinity fractions, in order to calculate an average crystallinity fraction we
    perform a number density weighted average out to the temperature cut-off.

\section{Images}\label{sec:images}
\begin{figure*}
\centering
\includegraphics[width=\textwidth]{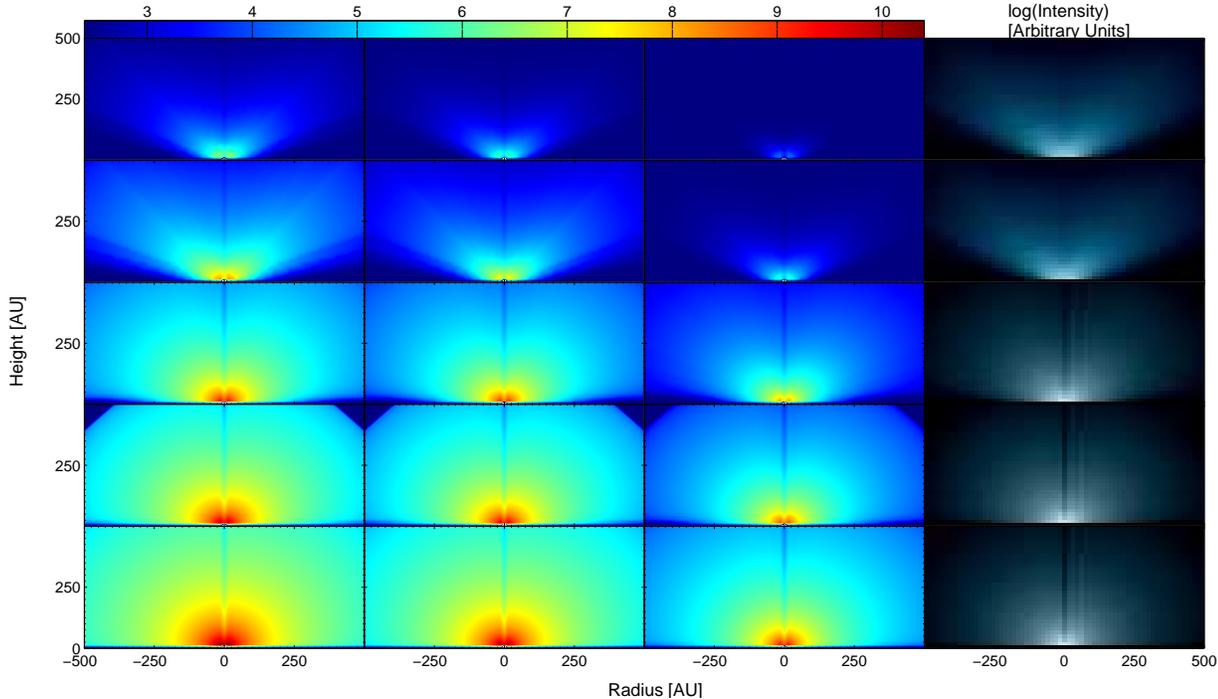}
\caption{ Synthetic images for disc models with irradiating fluxes 
    $\Phi=10^{41},10^{42},10^{43},10^{44},10^{45}$erg s$^{-1}$,
   the far left hand column shows the image in the H band, the next
   column displays the K band and the next column the L band. The far right
    column displays an RGB composite image (L,K \& H bands respectively). The images are individually scaled
    so that there is a 5dex spread between the brightest pixel and the
    darkest. All images assume the disc is edge-on (i.e. an inclination of 90$^o$), therefore we block out the star assuming it would be blocked out by the presence of an optically thick disc.}\label{fig:images}
\end{figure*}

In Figure~\ref{fig:images} we show calculated images of the 5
    different models in the H, K' \& L'  bands, they all show a `wingnut'
    morphology. The emission is strongest in the bluer
    bands with the H band being strongest and the L' being the
    weakest.

    As the mass-loss rate increases the emission becomes more
    extended and reaches larger radii, since the density along a streamline falls to a given value at larger radius. We also find that all of the
    emission is scattered light from the central star and inner disc,
    many orders of magnitude above any thermal emission from the
    wind itself, in agreement with our assertion that the wind is optically
    thin. The fact that the emission is dominated by scattered light allows us
    to understand the evolution of the color variation and emission
    morphology as the mass-loss rates change. For high mass-loss rates
    the emission morphology follows the gas density distribution of
    the wind (i.e. essentially spherical with a gap on axis in a region where 
    the wind is negligible),  while more pronounced `wingnut' profiles
    are clearly visible for lower values of mass-loss rate. This arises from the fact that
    scattering at a given wavelength is dominated by dust grains of
    comparable sizes. Given the bands we are interested in are in the
    range 1-5$\mu$m then when there is large spatial variation in
    grains of this size then a profile that differs markedly from the
    gas density would be expected (e.g. the wingnut). However, for large
    enough mass-loss rates dust grains with sizes in the 1-5$\mu$m
    range can be entrained 
    in the flow everywhere and the  emission then simply follows the
    gas density distribution. We find that all of the synthetic
    images are dominated by blue light; this is somewhat expected
    since the smallest grains can always be entrained in the flow and
    outnumber the larger grains, thus dominating the opacity for an
    MRN size distribution as the one considered here.
  Furthermore, the negative slope with wavelength of the irradiating
  field, which in the 1-5$\mu$m range is dominated by emission from
  the disc, means that short wavelength photons are more abundant,
  contributing to the blue appearance of the synthetic images.
  
\begin{table}
\caption{Model Luminosities}
\centering
\begin{tabular}{c c c c}
\hline\hline
$\log(\Phi)$ s$^{-1}$ & & log(Luminosity) erg s$^{-1}$ Hz$^{-1}$ & \\
 &  H & K' & L' \\ \hline
41 & 15.8 & 15.4 & 14.0 \\
42 & 17.4 & 17.2 & 16.2 \\
43 & 18.3 & 18.2 & 17.5 \\
44 & 18.9 & 18.8 & 18.2 \\
45 & 19.4 & 19.3 & 18.7 \\
\hline
\end{tabular}
\label{table:table}
\end{table}
\section{Crystallinity Profile}\label{sec:cryst}
We have used the wind profiles and dust temperature information
    from the radiative transfer calculation
    (Section~\ref{sec:rad}) to calculate crystallinity fractions in
    the wind. We have assessed the impact of the wind on the
    observable crystallinity fraction for discs, where the inner disc
    is both obscured (edge-on) and unobscured. In the unobscured case,
    to observe the disc's crystallinity fraction the wind will also
    make a contribution along the line of sight. We compare the disc's
    emission to the wind's emission at 10$\mu$m and find that at all radii the disc dominates over the wind, meaning that the observed crystallinity fraction for non-edge on discs is unaffected by the photoevaporative wind. However, when the disc is observed close to edge-on the disc's contribution will be undetectable and the crystallinity fraction in the wind maybe studied.  

\begin{figure}
\centering
\includegraphics[width=\columnwidth]{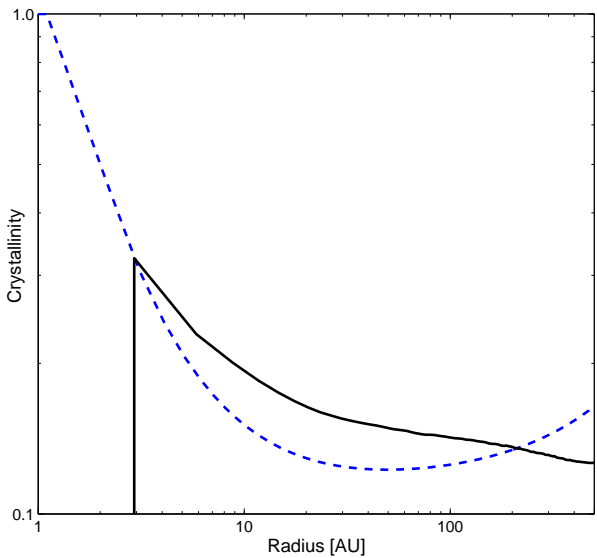}
\caption{Radial profiles of the wind crystallinity fraction (solid line) 
    calculated as a number density weighted mean
    in the wind and the underlying disc distribution taken
    from Dullemond et al. (2006) for an evolved 3.3Myr 2.5M$_\odot$ Herbig
    star (dashed line).}\label{fig:cryst}
\end{figure}

    It is important to note at this point, that the radial distribution of
    the crystallinity fraction in a wind is likely to differ
    from the disc's `native' distribution. In Figure~\ref{fig:cryst} we
    show radial profiles of the crystallinity fraction obtained from a
    photoevaporative wind  (solid black line) and  compare it to the input
    crystallinity distribution of the disc (dashed blue line).  The
  figure shows that at small radii $<2.5$AU 
    there is no crystallinity in the wind simply because there is no
    wind inside this radius to carry the dust grains to large
    heights.  Between 2.5-200AU there is a falling crystallinity
    fraction; however, it is always in excess of the disc's underlying
    fraction, whereas outside $200$AU it continues to fall even though the
    crystallinity fraction in the disc has now risen above the level in
    the wind. The crystallinity fraction in the wind remains the
    same for all ionizing luminosities and is insensitive to the
    temperature cut-off described in Section~\ref{sec:mcryst} in the
    range 150-50K. 

     The observed wind crystallinity profile is a direct
      consequence of the streamline topology, which implies that the
    grains in the wind at any given cylindrical radius must have originated
    in the disc at a smaller radius and hence, in general, with a higher crystallinity
    fraction. The rise in the disc crystallinity at
      a radii $>100$AU is not translated into the wind population, since most
    of the dust in the wind is entrained inside
    $5r_g\approx100$AU. The low wind rates at $>100$AU, coupled
    with the geometric dilution of the wind, mean that the crystallinity fraction in
    the wind continues to fall with radius. The fact that the
    crystallinity fraction is insensitive to the mass-loss rate is simply
    a consequence that the number of dust particles at a given point
    scales approximately as $n_{\textrm{dust}}\propto\Phi^{1/2}$ everywhere, the
    same as the gas, since it is largely unaffected by the change in
    maximum grain size for the MRN dust distribution chosen.  

\section{Comparisons to Observed Edge-on Objects}\label{sec:compare}
There are several objects that show extended emission above and below
    the dark dust lane indicative of an optically thick disc. All the
    objects in the sample presented by Padgett et al. (1999), which
    includes the well known
    `Butterfly Star', show extended
    scattered light emission above and below a dark dust lane. 
      The objects in  the Padgett et al. (1999) sample; however, all are 
    relatively low-mass young stars and the high fluxes detected
    in scattered light indicate that there is much more mass present
    than would be expected in a photoevaporative wind.
      Furthermore the observed structure is much more filamentary
    than our models predict. These observations clearly indicate that a photoevaporative wind is too weak to have produced the observed extended emission. Padgett et al. (1999) suggested that these objects are young (Class I) objects, with the extended emission arising from an infalling envelope. Wolf et al. (2003) and Stark et al. (2006) have successfully reproduced these observations using detailed radiative transfer modelling that include the combination of a disc component, an in-falling envelope and an outflow cavity. 

    PDS 144N (Perrin et al. 2006); however,  is a more promising
    candidate for a photoevaporative wind, with images showing a
    morphology very similar to the `wingnut' shapes predicted for the
    lower mass-loss rates $\Phi\le 10^{43}$erg s$^{-1}$. Furthermore adopting the $\Phi=10^{43}$erg s$^{-1}$, which predicts a surface brightness of 0.01~Jy~arcsec$^{-2}$ in H at a radius and height of $(R=100,z=100)$AU, we can derive a distance to PDS 144N. By comparing the observed image to our predicted image we find a distance of 200-400pc, which is in the range of previous estimates\footnote{Previous estimates range from 140 to 2000pc. Perrin et al. (2006) use a nearby A5 Herbig star (PDS 144S) to derive a distance of 1000pc, taking the system to be a wide binary. However, such a distance  places the system ~300pc above the Galactic plane. This would require the ejection of the wide binary system from a dense cluster, or formation in situ, both of which are unlikely.}.   As discussed
    in Perrin et al. (2006), it is highly 
    unlikely that the morphology arises from foreground extinction as
    suggested explanation to the morphology of the Padgett et
    al. (1999) sample.
    Perrin et al. (2006)  indeed discuss photoevaporation as a
    possible source for the extended emission. However, they  oversimplify the
    photoevaporation model by just considering the radial scale
    $r_g$, ignoring the
    fact that, as discussed in Section~2, dust grains can be carried to
    very large distances in the wind, and that at several flow scale
    heights from the disc the streamlines are spherical. Our models
    show that the use of $r_g$ in discussing the characteristic scales
    of the flow in dust is a poor approximation, since, as shown in
  Figure~\ref{fig:amax}, dust entrained at 
    $\sim1r_g$ can be easily carried to radii and heights of
    $>20r_g$ producing a dust density that varies on scales different
    to $r_g$ as shown in Figure~3. Perrin et al. (2006) present models of the object in which they
    construct an in-falling envelope with a jet cavity surrounding an
    extended passive accretion disc. While their model
    (Figure~6 of Perrin et al. 2006) can also reproduce extended
    emission,  it cannot reproduce the distinctive `wingnut'
    morphology seen in the observations and in the model presented in
    this work (see the top panels of Figure~\ref{fig:images}).        

\begin{figure}
\centering
\includegraphics[width=\columnwidth]{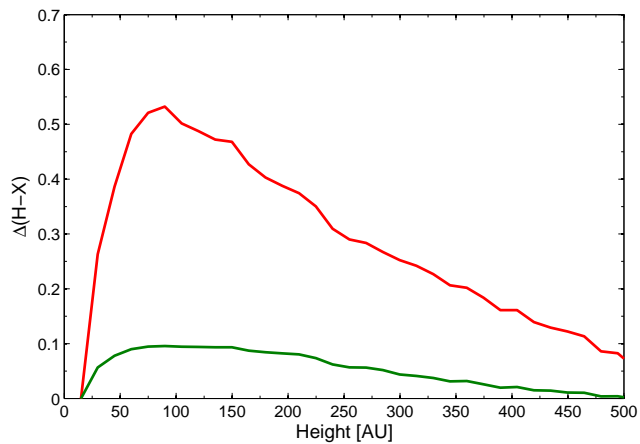}
\caption{Change in colour ($\Delta$H-X) as a function of height
above the midplane, at a cylindrical radius of 100AU. Green line:
    X=K' band colour; red line: X=L' band colour. The model shown is
    the $\Phi=10^{43}$ s$^{-1}$ calculation.}\label{fig:colour}
\end{figure}

    However both the model presented by Perrin et al. (2006) and the ones
    presented in this work fail to reproduce one important aspect of
    the observations. The observed colour of PDS 144N is
    such that the extended regions of the emission become dominated by
    the redder scattered light. This is opposite to what is
    predicted by the Perrin et al. (2006) model which predicts a colour
    variation of red to blue with height and the blue band dominating
    the emission at large height. This is due to a dust population whose maximum
    grains size decreases with height, as you move from the disc
    population to the envelope (ISM like) population. On the contrary,
    our model {\it can} reproduce the sign of the observed colour
    variation i.e that the relative strength of the red light increases
    with height. This is demonstrated in Figure~\ref{fig:colour} where we
    plot the change in H-K' and H-L' for the
    $\Phi=10^{43}$ s$^{-1}$ model as a function of height above the
    mid-plane for the $\Phi=10^{43}$ s$^{-1}$, where there is variations
in the grain population at sizes of a few microns, which dominate the
scattering opacity at NIR wavelengths. While our models predict that
the colours become redder with height above the disc (due to the fact
    that larger grains are entrained at greater height, see
    Figure~\ref{fig:amax}, and these scatter red light more efficiently), our models predict
scattered blue light to remain dominant at large heights (see
Figure 4). As discussed in Section~\ref{sec:images}, in our
      model blue light dominates due to the spectral slope of the
      irradiating SED and the fact that small grains are present in
    the wind at all heights.  We estimate using
    Figure~3 of Perrin et al. (2006) a change in H-L' ($\Delta$(H-L)) of $>$1.5, at least
    three times
    larger than our simple  model can predict.    
However, in 
    order to allow the redder bands to dominate over the bluer bands
    one needs to (i) increase the available disc emission at the
    redder wavelengths so there is more original flux to scatter and/or
   (ii) remove the smaller grains which are currently dominating
   the scattered light (iii) increase the optical depth to the star/disc so the extended region becomes optically thick, resulting in the reddening of photons as they pass through.  Grain growth in the disc provides a natural
 solution to both (i) and (ii).  Dullemond \& Dominik (2005) have
   shown that the
 dust grains in a disc easily grow to $\sim 1\mu$m reducing the
 opacity of the disc at shorter wavelengths and 
    resulting in a disc spectrum that falls less steeply or even rises
    to longer wavelengths in the 1-5$\mu$m range. Furthermore, the
  removal of small grains from the disc will also result in the removal of small
    grains from the wind, reducing the scattering efficiency at
      the shorter bluer radiation.  

Future observations of
    the $10\mu$m silicate feature  may help disentangle the origin of the
    dust particles in the extended region around PDS 144N since an
    in-falling envelope would be entirely composed of amorphous ISM type
    grains, while, as shown by our models, a photoevaporative wind
    origin implies the presence of crystalline silicate grains in the
    extended regions.  

\section{Conclusions}\label{sec:conclusions}

We have considered the fate of dust grains that can be entrained in a
    photoevaporative wind from the surface of a disc surrounding a
    Herbig Ae/Be star. For a median mass-loss rate of
    $\sim10^{-9}$M$_\odot$yr$^{-1}$, we find that grains up to  radii of
    several microns can be entrained in the wind. We also
    show that, once entrained in the wind, the dust grains will remain
    entrained and be carried out to very large radius. 

    We have considered the observational imprint of this
    wind-entrained dust on edge-on discs, showing 
    that the combination of a photoevaporative wind structure and a variable
    dust grain population resulting from the variable drag force at
    the base of the wind,  can naturally reproduce a `wingnut'
     morphology of the dust density distribution in the
      wind. Using a combination of Monte-Carlo and ray 
    tracing radiative transfer techniques, we calculate scattered light
    images from the density distribution at near infared
    wavelengths.  We find our model is not applicable to the well known edge-on discs in the Padgett et al. (1999) sample, which are too young and optically thick to be explained  with a photoevaporative origin, but are likely to arise from an infalling envelope as shown in Wolf et al. (2003) and Stark et al. (2006). These synthetic images show a wingnut
    morphology inferred from the dust density distributions, similar to
    observations of the edge-on disc around PDS 144N (Perrin et
    al. 2006).The synthetic images, however, are dominated by emission from the smallest
    grains entrained in the flow, hence failing to reproduce the
    colour gradient of the observations, which show redder emission at
    larger heights above the disc. Grain growth in the disc, is a
    natural solution to the colour problem, and we estimate
    severely depleted abundance of grains with radii smaller than
    $1\mu$m.

  Finally we consider the crystallinity of dust grains entrained in the
    flow: By following the thermal evolution of the grains, we find
    that crystallinity fraction will remain unchanged when the disc's
    photosphere can be observed since the observable mass in the wind
    in much less than the observable mass in the disc. However when
    the disc photosphere is obscured (i.e. for edge-on discs) the
    crystallinity fraction in the wind is significantly enhanced 
    above the discs underlying crystallinity fraction.  Finally, we
    suggest that detection of crystalline
    grains in extended emission around an edge on disc is indicative
    of a photoevaporative wind and argues against an envelope origin
    for the extended emission.

\section*{ACKNOWLEDGMENTS}
We wish to thank the referee, Barbara Whitney for her help with improving this paper. 
JEO acknowledges support of a STFC PhD studentship an is indebted to
    the University of Exeter Astrophysics Department for hospitality during the completion of
    this work. BE is supported by a Science and Technology Facility Council Advanced Fellowship. Some of this work was performed using the Darwin Supercomputer of the University of Cambridge High Performance Computing Service (http://www.hpc.cam.ac.uk/), provided by Dell Inc. using Strategic Research Infrastructure Funding from the Higher Education Funding Council for England.

\label{lastpage}

\end{document}